\documentclass[a4paper,11pt]{article}
\usepackage{aaskaiid}
\newcommand{\apjl}{ApJL} 
\newcommand{\apjs}{ApJSS} 
\newcommand{\ao}{Appl. Opt.} 
\newcommand{\aap}{A\&A} 

\usepackage{aas-macros}
\usepackage{orcidlink}

\title{Role of SKA in Advancing Remote Measurements of Magnetic Fields of Solar Coronal Mass Ejections}
\ShortTitle{Remote sensing CME magnetic fields}

\author[1,2,3]{Devojyoti Kansabanik\orcidlink{0000-0001-8801-9635}}
\ShortName{Kansabanik et al.}
\author[4]{Surajit Mondal\orcidlink{0000-0002-2325-5298}}
\author[6,7]{Shaheda Begum Shaik\orcidlink{0000-0002-3089-3431}}
\author[5,3]{Peijin Zhang\orcidlink{0000-0001-6855-5799}}
\author[4]{Puja Majee \orcidlink{0000-0002-2711-2366}}
\author[2]{Angelos Vourlidas\orcidlink{0000-0002-8164-5948}}
\author[8]{John Morgan\orcidlink{0000-0001-9224-5483}}
\author[4]{Divya Oberoi\orcidlink{0000-0002-4768-9058}}
\author[9]{Anshu Kumari\orcidlink{0000-0001-5742-9033}} 
\author[10]{Alec Thomson\orcidlink{0000-0001-9472-041X}}
\author[5]{Bin Chen\orcidlink{0000-0002-0660-3350}}
\author[11]{Carl Shneider\orcidlink{0000-0002-3689-6959}}
\author[10]{Hariharan Krishnan} 
\author[12]{Kamen Kozarev\orcidlink{0000-0002-6591-4482}}
\author[13]{Sneha Pandit\orcidlink{0000-0002-9921-939X}} 
\author[14]{Vanessa Moss\orcidlink{0000-0002-3005-9738}}

\affiliation[1]{Instituto de Astrofísica de Andalucía-CSIC (IAA-CSIC), Glorieta de la Astronomía s/n, E-18008, Granada, Spain}
\affiliation[2]{Johns Hopkins University Applied Physics Laboratory, 11001 Johns Hopkins Rd, Laurel, MD, USA}
\affiliation[3]{University Corporation for Atmospheric Research, 3090 Center Green Dr., Boulder, CO, USA}
\emailAdd{devojyoti96@gmail.com}
\affiliation[4]{National Center for Radio Astrophysics, Tata Institute of Fundamental Research, S. P. Pune University Campus, Ganeshkhind, Pune, India}
\affiliation[5]{Center for Solar Terrestrial Research, New Jersey Institute of Technology, NJ, USA}
\affiliation[6]{George Mason University, Fairfax, VA, USA}
\affiliation[7]{US Naval Research Laboratory, Washington,  DC, USA}
\affiliation[8]{CSIRO Space and Astronomy, Bentley, WA, Australia}
\affiliation[9]{Udaipur Solar Observatory, Physical Research Laboratory, Udaipur, India}
\affiliation[10]{Square Kilometre Array Observatory, Jodrell Bank, Lower Withington, Macclesfield Cheshire, SK11 9FT, United Kingdom}
\affiliation[11]{SnT (Interdisciplinary Centre for Security, Reliability, and Trust), University of Luxembourg, Luxembourg}
\affiliation[12]{Institute of Astronomy, Bulgarian Academy of Sciences, Bulgaria}
\affiliation[13]{Inter-University Center for Astronomy and Astrophysics, Pune, India}
\affiliation[14]{CSIRO Space and Astronomy, Marsfield, NSW, Australia}

\abstract{Coronal Mass Ejections (CMEs) are large expulsions of magnetized plasma from the Sun into interplanetary space and are the primary drivers of extreme space weather variations. The strength and topology of CME magnetic fields largely determine their impact on Earth. Although visible-light coronagraphs routinely observe CMEs and provide their geometric and kinematic properties, they cannot directly measure CME vector magnetic fields. These fields evolve from initiation through the inner heliosphere due to interactions with other CMEs, coronal structures, and the ambient solar wind, leading to significant structural deformation. Such evolution complicates predictions of the CME magnetic field at Earth. Accurate measurements of CME magnetic fields in the corona and heliosphere are therefore essential for advancing space weather forecasting. Radio observations spanning MHz to GHz frequencies provide a powerful remote-sensing approach for measuring CME magnetic fields from the ground. Recent observations with Square Kilometre Array (SKA) precursors and pathfinder instruments, as well as other new-generation facilities, have demonstrated the potential of these radio techniques for CME magnetic-field diagnostics. At the same time, these studies have highlighted several limitations of current instruments. The higher sensitivity, wider instantaneous bandwidth, and broader frequency coverage of the SKA will open a new observational window, enabling these techniques to be fully exploited for constraining SpWx models and improving predictive accuracy. However, such observations are non-standard and require special consideration in scheduling, calibration, and imaging. Developments achieved with SKA precursors and pathfinders are paving the way for robust CME magnetic-field measurements with the SKA.}

\begin{document}
\maketitle

\section{Introduction}\label{sec:intro}
Coronal mass ejections (CMEs) are massive eruptions of magnetized plasma from the solar corona into the heliosphere \citep{Chen2011, Webb2012}. CMEs have major societal relevance, as they drive the most severe space weather events capable of disrupting technological and space-based systems, such as satellites, power grids, and communication \citep{Owens2021, Cliver2022}, popularly known as the geo-effectiveness of the CME. Magnetic field of CME is a crucial component not only determining its space weather importance, but it is fundamental to understanding coronal magnetic field evolution, helicity transport, eruption mechanisms \citep{Low1996,Low2001,atmos10080468,Foullon_2013,Zhelyazkov2015,Vashalomidze2022}, and particle acceleration processes \citep{Mikić2006,Vourlidas2012,Xia_2020,Frassati_2022}.

Since the discovery of CMEs using space-based coronagraph observations \citep{Tousey1973, Koomen1975}, extensive remote-sensing and in-situ observations have advanced our knowledge about CME initiation, structure, propagation, and geo-effectiveness \citep[e.g.,][]{rodriguez2006, Rodriguez2011, Mishra2021, Davies2021}. Yet, several fundamental questions remain unresolved. The magnetic configuration in the corona evolves until it reaches a critical instability and erupts, but the precise triggering mechanism of the eruption is still debated \citep{MITTAL2010, Georgoulis2019}. Similarly, the nature of early acceleration of CMEs \citep{JamesChen2003, SURYANARAYANA20191} and the role of magnetic reconnection in driving this process \citep{Wu2005} are not fully constrained. Measuring the magnetic field in the solar corona, particularly within CMEs, remains one of the most persistent challenges in heliophysics. The difficulty arises from a combination of physical, instrumental, and geometric factors that limit both direct and indirect magnetic diagnostics of CMEs across different heliocentric heights.

\begin{figure*}[!htbp]
    \centering
    \includegraphics[trim={1cm 1.5cm 0cm 4cm},clip,scale=0.5]{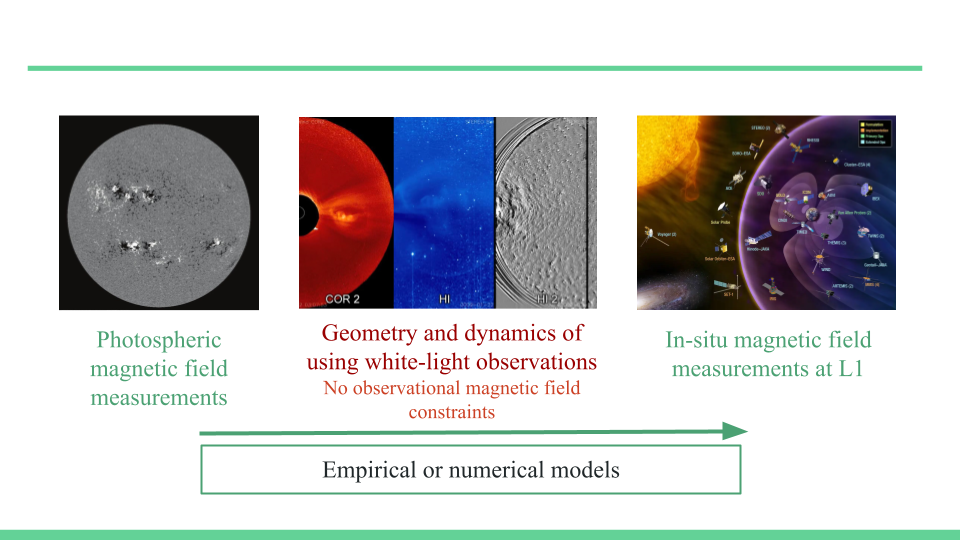}
    \caption{A schematic depicts the current approach of CME magnetic field prediction for space weather forecasting. The left panel shows the photospheric magnetic field measurements, the middle panel shows white-light observations used to constrain geometric and dynamical properties, and the right panel shows the swarm of spacecraft used for in-situ measurements.}
    \label{fig:cur_approach}
\end{figure*}

Kinematic properties of CMEs, such as speed, acceleration, and mass, are routinely measured using white-light coronagraphs. Their speeds range from a few tens to several thousand km/s, allowing some to reach Earth within hours to days. However, beyond the dynamical and thermodynamic properties, the magnetic field of CMEs remains the key physical parameter governing CME initiation, evolution, and impact. In-situ measurements provide magnetic field data, but only after CMEs have propagated far from the Sun, expanded significantly, and interacted with the background solar wind. In-situ spacecraft offer snapshots of CME magnetic fields at their encounter points. Still, these measurements represent only localized cuts through structures that have evolved for tens to hundreds of solar radii. Multi-spacecraft encounters remain rare, and even in such cases, the spatial sampling is too sparse to capture the global magnetic structure of CMEs.

Current space weather prediction, as depicted in Figure \ref{fig:cur_approach}, uses only photospheric magnetic field measurements (left panel) and in-situ measurements at the Sun-Earth L1 point (right panel). Between them in the corona and heliosphere, magnetic field measurements are generally unavailable, where empirical or magnetohydrodynamic (MHD) models are constrained by photospheric magnetic field measurements and by geometric and dynamic properties derived from white-light observations (middle panel) to infer magnetic fields.  

The southward component of the CME magnetic field ($B_\mathrm{z}$) with respect to Earth's magnetic field is particularly critical, as it reconnects with Earth’s northward magnetic field, driving geomagnetic storms and energetic particle injections. Accurate forecasting of CME geo-effectiveness, therefore, requires reliable estimation of CME-$B_\mathrm{z}$, yet current forecasts remain limited by sparse vantage points, complex CME solar-wind interactions, and the vast spatial scale of CMEs relative to Earth. These CME sizes at 1 AU are enormous, typically ranging from 0.03 AU to 1.34 AU in radial extent, much larger than the Earth–Moon system \citep{Mishra2021_radial_size}. This immense scale complicates the attempts to model or reconstruct their 3D magnetic configuration and evolution accurately. 

Forecasting CME geo-effectiveness and impacts thus requires a two-step approach: (i) reconstructing the 3D magnetic field near the Sun, and (ii) tracking its evolution through the corona and heliosphere through 1 AU. Because CME magnetic fields evolve through interactions with ambient solar wind and large-scale heliospheric structures, improving physics-based models and constraining them with multi-height magnetic field observations are essential for reliable CME-$B_\mathrm{z}$ predictions. 

This chapter focuses on different remote sensing techniques in radio wavebands in probing CME magnetic fields, a domain that has gained renewed attention thanks to recent advances in radio interferometry. Various direct and indirect radio observables have demonstrated the potential to infer CME magnetic field strengths and configurations \citep{chen2025measuring}. SKA precursors and pathfinders have already demonstrated these valuable capabilities \citep[e.g.,][]{Mondal2020a,Kansabanik2023_CME1}, representing a significant leap over the capabilities of earlier radio arrays. However, even these instruments remain limited in terms of sensitivity, frequency coverage, dynamic range, and temporal imaging cadence, preventing the full exploitation of the potential of radio diagnostics for CME magnetic field measurements. 

SKA telescopes are poised to overcome these limitations with their combination of wide instantaneous bandwidths, high sensitivity, and high-fidelity snapshot imaging capability, providing the SKA with the capabilities for detailed mapping of CME B-structures. It will bridge the critical observational gap between the low corona and interplanetary space, where CMEs undergo the most dynamic magnetic evolution. Through these advances, SKA promises to transform our ability to remotely measure and model CME magnetic fields, thereby advancing both the scientific understanding of CME science and the practical applications in the prediction of their space weather consequences.

\begin{figure*}
    \includegraphics[trim={1.7cm 2cm 2cm 3cm},clip,scale=0.5]{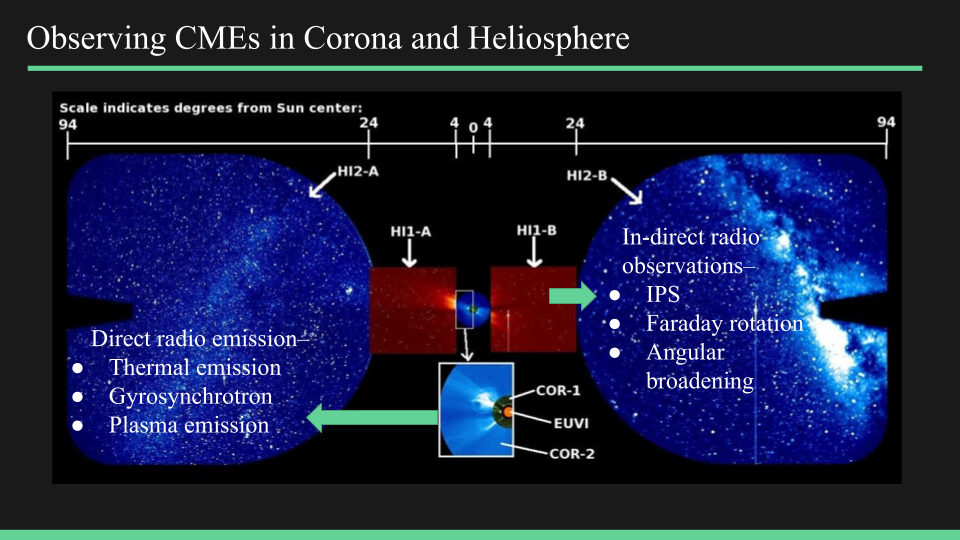}
    \caption{Direct and indirect radio observables that enable measurements of CME magnetic fields in the corona and inner heliosphere using ground-based radio observations. The background illustration highlights the imaging capabilities provided by the combined instrument suites onboard the STEREO-A (Ahead) and STEREO-B (Behind) spacecraft. The central panel presents a composite image that merges extreme ultraviolet (EUV) and white-light coronagraph data, incorporating observations from the EUVI imager along with the COR1 and COR2 coronagraphs. Additionally, the HI1 and HI2 white-light heliospheric imagers extend this capability by enabling large-scale imaging of the heliosphere. The figure is adapted from \citep{Vourlidas2021}.}
    \label{fig:radio_observables}
\end{figure*}

\section{Radio Diagnostics of Remote Sensing CME Magnetic Fields}\label{sec:radio_diag}
Different parts of CMEs emit radio emissions through different emission mechanisms at various stages of their evolution. All of these emissions carry imprints of the magnetic field and provide remote-sensing tools to probe CME magnetic fields, which are otherwise unavailable \citep{Vourlidas2020,Carley2020}. Beyond these direct radio emissions from CME plasma, there are other indirect observing techniques at radio wavelengths using background galactic/extra-galactic radio sources that can also be used to study CME magnetic fields at higher coronal and heliospheric heights. Across the electromagnetic spectrum, only white-light and radio observations can continuously probe CME properties from the low corona out into the heliosphere, as illustrated in Figure \ref{fig:radio_observables}. Moreover, the plasma parameters diagnosed by these two bands are complementary -- white-light is sensitive to density structures, while radio wavelength is sensitive to magnetic fields -- providing a more complete characterization of CME evolution. In the following Sections \ref{subsec:direct_radio_obs} and \ref{subsec:indirect_radio_obs}, we discuss these observing techniques at radio wavelengths, which can be used for remote measurements of magnetic fields in different parts of CMEs. 

\subsection{Direct Radio Emissions}\label{subsec:direct_radio_obs}
Direct radio emission from CME plasma can provide measurements of specific components or the magnetic fields. Detailed descriptions of these emission mechanisms are presented in another chapter \citep{Patra01.2026.SKA}. Recovering the full vector magnetic field, therefore, relies on 3D magnetic field models constrained by these observations. Although recent work has demonstrated the promise of these techniques, observational limitations remain -- limitations that the SKA telescopes are well equipped to overcome. The following subsections will provide a brief overview of the capabilities of different direct observables for probing CME magnetic fields in the corona. 

\begin{figure*}[!htbp]
    \centering
    \includegraphics[width=0.5\linewidth]{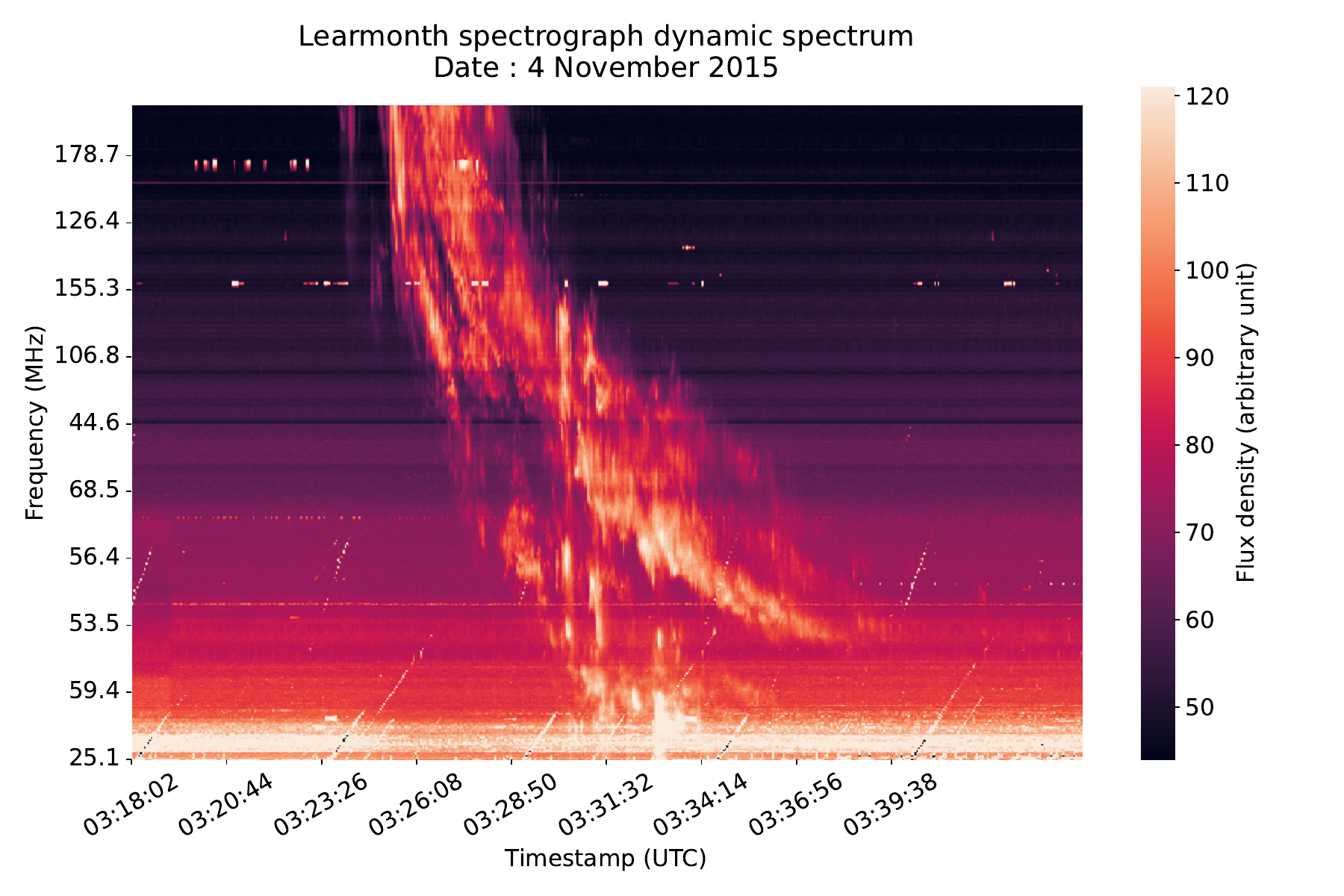}\includegraphics[width=0.5\linewidth]{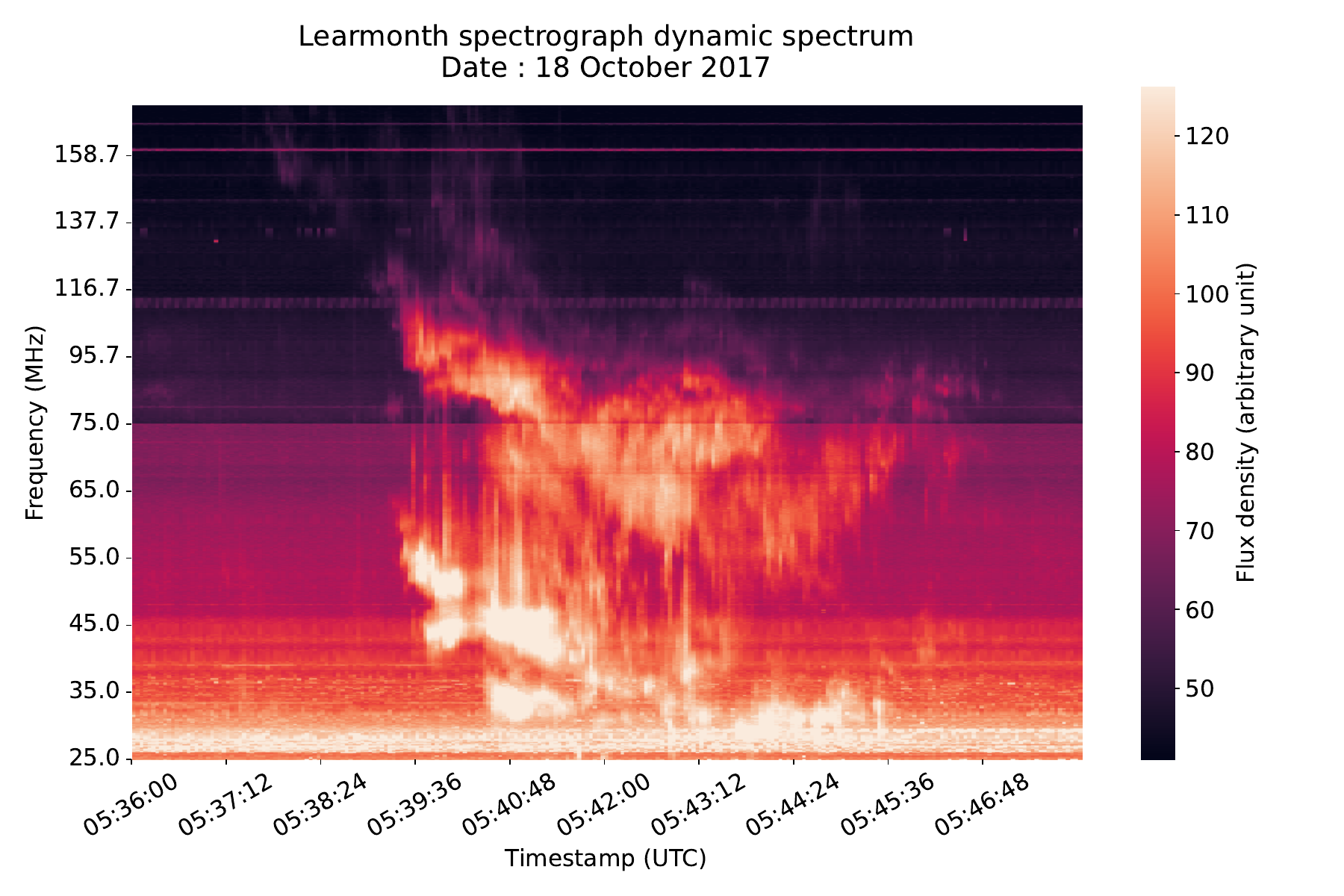}
    \caption{Left panel: A type-II radio burst showing multi-lane emission in the solar dynamic spectrum observed using the Learmonth solar radio spectrograph. Right panel: A type-IV radio burst observed with the Learmonth solar radio spectrograph.}
    \label{fig:typeII_IV}
\end{figure*}

\subsubsection{Magnetic Field at CME Shocks using Type-II Radio Bursts}
Type-II solar radio bursts (type-IIs) are produced by the coherent plasma emission mechanism when energetic non-thermal electrons are accelerated by the shock that is most commonly driven by CMEs. In the dynamic spectrum, type-IIs appear as slowly drifting ($\sim0.05\rm MHz/s$ at $\rm100MHz$) emissions as shown in the left panel of Figure \ref{fig:typeII_IV}, both at the fundamental and harmonic of the local plasma frequencies. Each of these emission lanes often exhibits multilane structures, which are interpreted either as emissions occurring from upstream and downstream of the shock \citep{Smerd1974, vrsnak2001}, or multi-location emission from different parts of the shock front \cite{zucca2025multilane,zhang2024imagingtype2}.

There are two known methods to infer the magnetic field at the coronal shock using type-II radio bursts: i) {\it Band-splitting of type-IIs} -- when fundamental and harmonic lanes themselves show two split-bands which are morphologically similar in dynamic spectra originating from upstream and downstream shocks. This band-splitting feature of type-IIs is used to estimate the strength of the shock-entrained magnetic field from the background Alfv\'en speed by combining the frequency drift rate (shock speed), frequency ratio of the split bands (density compression across the shock), and the electron density (source height) \citep[e.g.,][]{Smerd1975,Vasanth2014,Hariharan2015,Kumari2017typeII_band,MAHROUS201875}. While this approach is strongly dependent on the assumed electron density model and shock geometry, multi-wavelength observations, such as extreme ultraviolet (EUV) images and white-light coronagraph images, have been incorporated to reduce the model uncertainties \citep{Gopalswamy2012,Kumari2017typeII_band,Kumari2019}. ii) {\it Type-II harmonic emissions} -- exhibiting weak circular polarization, which is directly related to the magnetic field. Hence, this circular polarisation fraction provides estimates of magnetic field strength at the shock front of CMEs \citep[e.g.,][]{Hariharan2014,Kumari2017a,Ramesh_2022, Ramesh_and_Kathiravan2022, Ramesh_2023}.

\subsubsection{Magnetic Fields of CME Core using Type-IV Radio Bursts}
Type-IV radio bursts are long-duration, broadband, and relatively bright continuum features in dynamic spectra \citep{Boischot1957}. They are classified into moving and stationary types: moving bursts show frequency drift and are linked to outward-moving CME-related sources, whereas stationary bursts remain fixed and are associated with the CME wake. A sample moving type-IV burst is shown in the right panel of Figure \ref{fig:typeII_IV}. Although easily identifiable in dynamic spectra, their emission mechanisms vary from event to event \citep{morosan2019}. Some studies attribute them to gyrosynchrotron emission from mildly relativistic electrons in CME magnetic fields \citep[e.g.,][]{Boischot1968, Dulk1973}, while others found plasma emission as a suitable mechanism \citep[e.g.,][]{Weiss1963, Gary1985} due to their high brightness temperature and strongly polarized nature. The mechanism may also evolve during a single event \citep{morosan2019}. Spectropolarimetric modeling of gyrosynchrotron emission from type-IV bursts can be used to measure the magnetic field strengths of the CME cores. Similarly, the polarisation of plasma emission from type-IV bursts can also provide magnetic field strengths of CME cores. Hence, after identification of the emission mechanism, type-IV bursts have been used to measure magnetic fields of the CME core \citep[e.g.,][]{Raja2014, Bain2014, Carley2017}.

\subsubsection{Circular Polarization of Thermal Emission}
Thermal free-free emission is produced by free electrons present in the CME plasma. This emission has a much lower brightness temperature. Hence, only a handful of studies have claimed to detect thermal free-free radio emission from CMEs \citep{Gopalswamy1992,Gopalswamy1993,Ramesh2021}. Thermal free-free emission has been used to measure the mass of the CME \citep{Gopalswamy1992}. Thermal emission can show a small amount of circular polarization in the presence of a magnetic field in the plasma \citep{Sastry_2009}, which can be used to measure the line-of-sight magnetic field of the CME plasma \citep{Ramesh2021}. On the other hand, thermal electrons gyrating in the CME magnetic field may emit highly polarised gyroresonance emission at gyro-frequency or its harmonics. Based on high polarisation fraction, the gyro-frequency and its harmonics can be identified, which provide a direct measurement of the magnetic field strength of the emitting plasma. Gyroresonance emission is another type of thermal emission capable of providing CME magnetic field measurements \citep{mondal2025possible}. 

\subsubsection{Gyrosynchrotron Emission}
\begin{figure}
    \centering
    \includegraphics[trim={5.3cm 9cm 0cm 0cm},clip,width=0.6\linewidth]{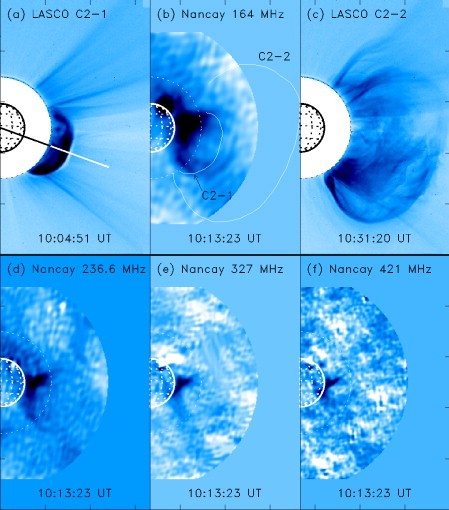}
    \caption{The first detection of gyrosynchrotron emission from CME plasma using Nan\c{c}ay Radio Heliograph. The left panel shows the gyrosynchrotron radio emission from the CME, with white arcs marking the white-light CME, which is shown in the right panel. The right panel shows the CME observed at the closest time using the white-light LASCO C2 coronagraph (Reproduced from \citet{bastian2001}).}
    \label{fig:radio_cme_bastian}
\end{figure}

Gyrosynchrotron emission is produced by mildly relativistic electrons trapped in CME magnetic fields. While in some instances type-IV radio bursts are found to be produced by gyrosynchrotron emission, there is another type of gyrosynchrotron emission that has a morphology similar to the CME structures seen in white-light images. These emissions are generally faint and were first detected by \cite{bastian2001} (Figure \ref{fig:radio_cme_bastian}), who gave them the name ``radio CME". This was followed by only a handful of other successful imaging detections \citep{Maia2007,Mondal2020a}, where radio emissions trace the white-light morphology. While there is no consensus yet on what types of gyrosynchrotron emissions associated with CMEs should be referred to as ``radio CME" \citep{Vourlidas2020}, spectral modeling of gyrosynchrotron emissions from different parts of a CME does provide a unique observational tool to estimate CME magnetic field and other plasma parameters using the fast gyrosynchrotron codes \citep{Fleishman_2010,Kuznetsov_2021}.    

\subsection{CME Magnetic Field Measurements using Indirect Radio Observables}\label{subsec:indirect_radio_obs}
Direct radio emission from CMEs becomes fainter at higher coronal heights, and the characteristic frequencies also fall below the Earth's ionospheric cutoff ($\sim$10 MHz), making them difficult to observe using ground-based radio instruments. Since the magnetized plasma of the CME also affects the radio wave, which propagates through it, it allows background radio sources to be used as probes of the CME magnetic field in the outer corona and inner heliosphere.

\subsubsection{Interplanetary Scintillation}
Plasma density irregularities in the turbulent solar wind and CMEs cause intensity fluctuations in compact background radio sources, known as interplanetary scintillation (IPS; \citet{Clarke64}). By measuring these rapid variations -- produced as density structures drift across the line of sight -- and analyzing their strength and cross-correlation between antennas, IPS provides estimates of solar wind speed, turbulence levels, and large-scale flow structure. Multiple line-of-sight observations further enable three-dimensional reconstruction of solar wind streams and tracking of CMEs in the inner heliosphere. A detailed explanation of the IPS technique for measuring heliospheric plasma properties is provided in another chapter \citep{Chhetri01.2026.SKA}.

Over recent decades, increasingly advanced techniques have enabled 3D reconstruction of heliospheric density and velocity from multi-station IPS observations covering the whole heliosphere \citep[e.g.,][]{Jackson1997, Jackson1998}. While IPS does not directly measure magnetic fields, it is sensitive to the axial ratio of the turbulence (in the plane of sky), which is in turn dependent on the magnetic field (albeit with a 180$^\circ$ ambiguity). \citet{2002GeoRL..29.1913L} have shown how the IPS is sensitive to the swing of the magnetic field from parallel to perpendicular to velocity as a CME crosses the line of sight. Additionally, the velocity information from IPS can be used to accurately propagate solar magnetic fields out to 1 AU \citep{Jackson2020}. Additionally, IPS data, particularly those from global IPS networks, can be used in conjunction with heliospheric imager data and MHD models to model CMEs and their magnetic fields as they traverse interplanetary space \citep{Iwai2022}.

\subsubsection{Faraday Rotation Measurements}
When linearly polarized radiation propagates through a magnetized plasma, its polarization plane rotates—a phenomenon known as Faraday rotation \citep[FR;][]{Ferri2021}. The amount of rotation depends on the line-of-sight integral of the electron density multiplied by the line-of-sight magnetic-field component, and scales with the square of the observing wavelength. The wavelength-independent term is the Rotation Measure \citep[RM;][]{Brentjens2005}. FR measurements of background polarized sources seen through CMEs \citep[e.g.,][]{Bird-CME-FR-1985,kooi2017,Kooi2021} provide a powerful remote-sensing probe of CME magnetic fields in both the corona and inner heliosphere \citep[e.g.,][]{Vourlidas2020}. Recent progress in FR studies of CMEs is reviewed in \cite{Kooi2022}.

\subsubsection{Scatter Broadening}
Small-scale density fluctuations in the corona and solar wind scatter radio waves, causing angular broadening of compact sources \citep{Oberoi2023,zhang2025probingCrab}. A detailed description of this method is presented in other chapters \citep{Raja01.2026.SKA,PeijinZhang01.2026.SKA}. If the turbulence is anisotropic and elongated along the magnetic field, the scatter-broadened image becomes elongated perpendicular to the projected field direction. Measurements of the apparent axial ratio and position angle can therefore be used to infer the orientation of the CME magnetic field in the plane of the sky. For background sources seen through CMEs, temporal changes in the broadening pattern provide complementary constraints to FR, helping to reconstruct the 3D magnetic-field geometry of the CME plasma.

\subsection{Observational and modeling Requirements}
The radio-wavelength observational techniques discussed in Section \ref{sec:radio_diag} have already been shown to be feasible through demonstrations with earlier-generation instruments. However, the successful implementation of these observations depends on meeting several specific observational requirements. Both bright coherent emissions from CME shocks (type-II radio bursts) or cores (type-IV radio bursts) with brightness temperature go up to $\sim10^8$ K, and much fainter incoherent emissions from the CME plasma with brightness temperature $\sim10^4$ K, can occur simultaneously. To extract reliable magnetic field and plasma diagnostics from both, they must be detected simultaneously in the image plane, requiring high-dynamic-range spectropolarimetric imaging, which is now possible with modern ground-based radio telescopes. To constrain the CME magnetic field at larger coronal distances and into the inner heliosphere using background sources, simultaneous measurements along multiple lines of sight passing through the CME are required. This depends on instrumental sensitivity, field-of-view, and the distribution of suitable background sources. Moreover, as with all remote-sensing techniques, all of these radio techniques provide line-of-sight integrals and therefore require appropriate modeling to recover the true 3D CME structure.

Combining both direct and indirect methods using ground-based radio telescopes along with observations from coronagraphs and heliospheric imagers can provide remote measurements of CME plasma parameters, including its magnetic field. For the precise prediction of CME evolution, it is important to understand both coronal and heliospheric models of CMEs and constrain them well using observations. Although all of these observing methods seem promising, there are several challenges to overcome in terms of observation for each of them. 

\section{Recent Achievements in CME Magnetic Field Measurements }\label{sec:current_status}
Over the past decades, SKA precursor and pathfinder instruments have achieved significant progress in making the aforementioned radio techniques practically feasible for real-world applications. In this section, we briefly discuss these recent results, which prepared the path forward for the SKA.

\subsection{Detection of Faint Gyrosynchrotron Emission from CME}
After the first detection of gyrosynchrotron emission from a CME by \citet{bastian2001} (Figure \ref{fig:radio_cme_bastian}), till about 2020, all detections \citep{Tun2013,Bain2014,Carley2017} corresponded to extremely fast and energetic CMEs, with speeds exceeding about 1000 km/s. Due to this, it was unclear whether weaker CMEs can indeed produce such emissions and whether gyrosynchrotron emission can be used to measure CME magnetic fields, even for weak CMEs. The detections of gyrosynchrotron emission using SKA precursors (MWA and MeerKAT), even from weaker CMEs, make it evident that gyrosynchrotron emission from CME plasma can be routinely observed and exploited to measure CME magnetic fields in the lower and middle corona.

\subsubsection{Observations using the MWA at Low and Middle-corona}
\begin{figure}[!htbp]
    \centering
    \includegraphics[width=0.5\linewidth]{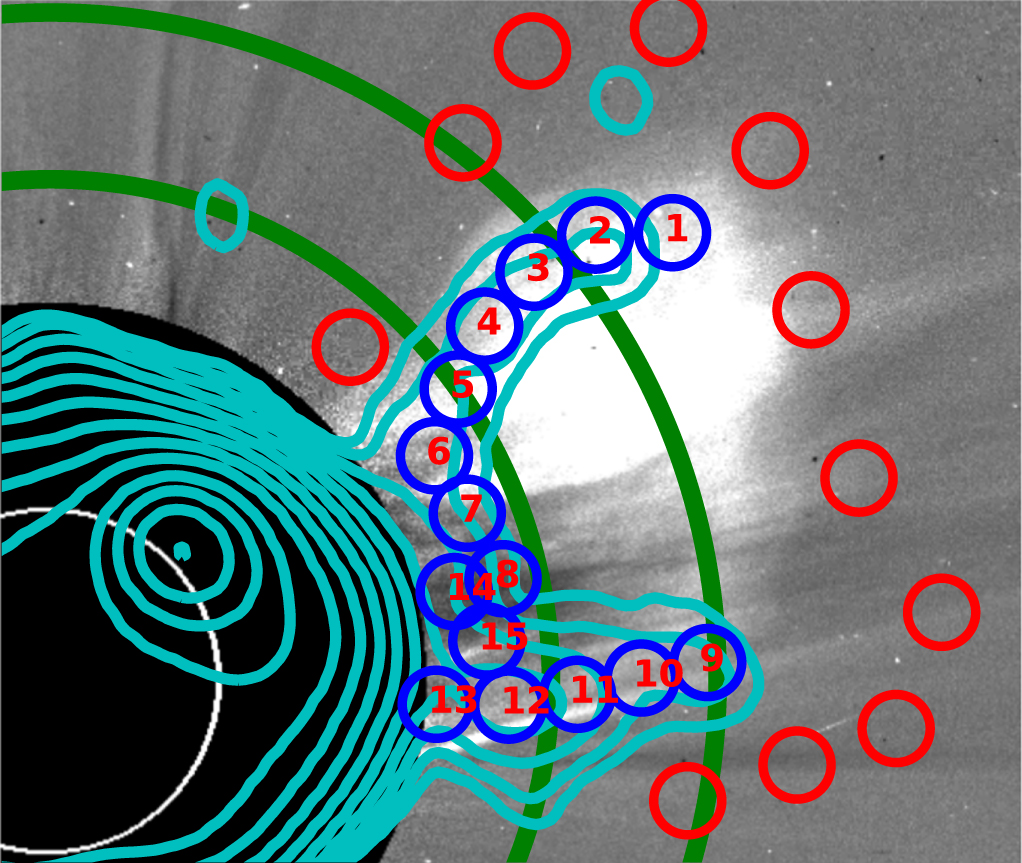}\includegraphics[trim={4.6cm 3cm 1cm 0cm},clip,width=0.44\linewidth]{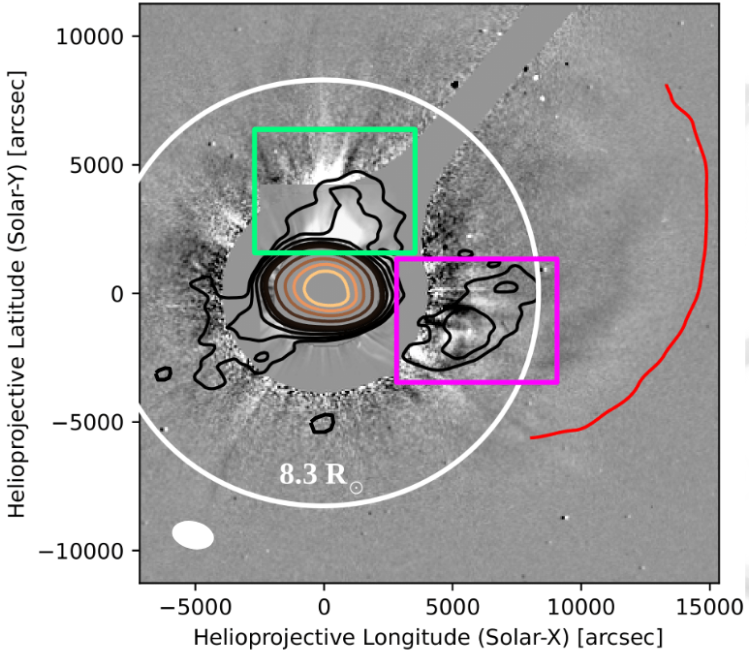}
    \caption{Left panel: The first detection of gyrosynchrotron emission from a slow CME. Cyan contours indicate the radio emission superimposed on the white-light LASCO C2 coronagraph image. Spectra have been extracted and modeled in the blue circle regions. (Reproduced from \citet{Mondal2020a}). Right panel: Gyrosynchrotron emissions detected from two CMEs, marked by green and magenta boxes, at the highest heliocentric distance of $\sim8.3\ R_\odot$. (Adapted from \citet{Kansabanik2023_CME1} and \cite{Kansabanik_2024}).}
    \label{fig:cme_gs}
\end{figure}

The first detection of gyrosynchrotron emission from a relatively weak and slow CME was reported by \citet{Mondal2020a} using the MWA, which is shown in the left panel of Figure \ref{fig:cme_gs}. The CME speed was only $\sim 400~\mathrm{km\,s^{-1}}$, significantly lower than previous cases. This provided the first indication that gyrosynchrotron emission can also originate from weak CMEs, and that magnetic-field estimates can be obtained regardless of CME strength, provided sufficient sensitivity and dynamic range are available with modern instruments. Using these high-dynamic-range images \citep{Kansabanik2022_paircarsI, Kansabanik_paircars_2} from the MWA observations, \cite{Kansabanik2023_CME1} and \cite{Kansabanik_2024} detected these faint gyrosynchrotron emissions from another two weak CMEs (shown in the right panel of Figure \ref{fig:cme_gs}) till $\sim$8.3 $R_\odot$ (right panel of Figure \ref{fig:cme_gs}). While these detections demonstrated that gyrosynchrotron emission is most likely produced by a large fraction of CMEs, including both fast and slow CMEs, they also exposed limitations in the modeling techniques. Gyrosynchrotron modeling without using polarisation information has degeneracies between several physical parameters. One of the crucial factors is the magnetic field strength and line-of-sight angle. \citet{Kansabanik2023_CME1} and \citet{Kansabanik_2024} showed that detailed polarisation measurements are essential for robust parameter determination and breaking these degeneracies obtained from modeling of the spectra. However, such measurements are currently unavailable for weak CMEs due to sensitivity limitations, which are expected to be overcome by the enhanced sensitivity of the SKA.

\subsubsection{Observation using MeerKAT at Low-corona}
\begin{figure}
    \centering
    \includegraphics[width=\linewidth]{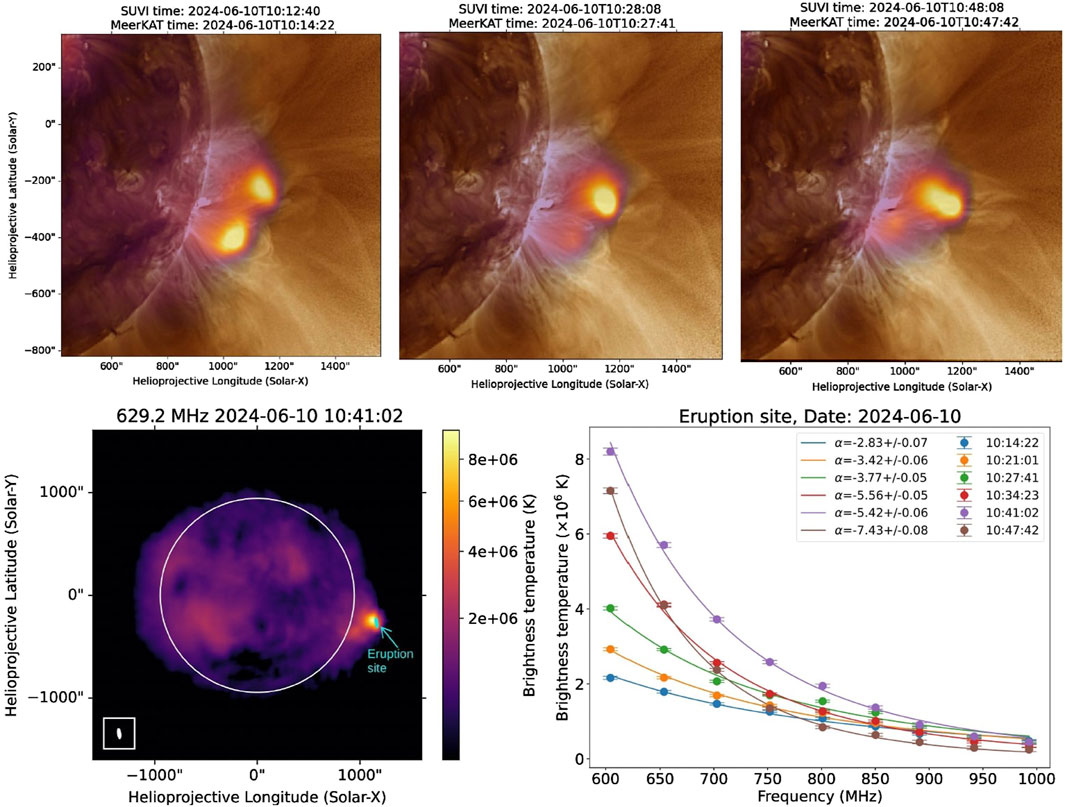}
    \caption{Science verification observation on 10 June 2024 using MeerKAT observed an erupting CME. The top panels show MeerKAT radio images overlaid on 195 Å EUV images. The bottom left panel displays the 629 MHz brightness temperature map during the CME, with a point-spread-function (PSF, shown in the bottom left corner inside the white marked box) sized region highlighted, while the bottom right panel shows spectra from this region at different times (Reproduced from \citet{Kansabanik2025_meerkat}).}
    \label{fig:cme_meerkat}
\end{figure}

MeerKAT \citep{meerkat2016} observations in the UHF (580–1015 MHz) and L-band (900–1670 MHz) can enable measurements of CME magnetic fields during the eruptive phase in the low corona ($\lesssim 2\,R_{\odot}$). Recently, \citet{Kansabanik2025_meerkat} demonstrated MeerKAT's solar observing capability. Science-verification observations on 10 June 2024 captured a CME eruption on the eastern limb as shown in the top panels of Figure \ref{fig:cme_meerkat}, revealing that the emission spectrum transitions from thermal to non-thermal gyrosynchrotron during the eruption as depicted in the bottom right panel of Figure \ref{fig:cme_meerkat}.

Thanks to its high sensitivity and excellent spectroscopic snapshot sampling in \textit{uv}-plane, MeerKAT produces high-dynamic-range solar images with quality comparable to observations at other wavelengths such as EUV. This marks the first demonstration with an SKA-mid precursor of detecting GS emission from CME plasma during eruption. Moreover, high-cadence (8s) spectroscopic snapshot imaging allows detailed tracking of the evolving CME magnetic field and other non-thermal plasma properties in the earliest phases of the eruption.

\subsubsection{Observations using the OVRO--LWA in Middle-corona}
The Owens Valley Radio Observatory Long Wavelength Array (OVRO–LWA) \citep{Gregg2023AASOVROLWA} is a low-frequency interferometer operating between $\sim$15 and 88 MHz. Its dense core, combined with baselines extending up to $\sim$2.5 km, provides the high surface-brightness sensitivity needed to detect gyrosynchrotron emission. Although OVRO–LWA is not formally an SKA precursor or pathfinder, its technical capabilities are comparable to those facilities.

\begin{figure}[!htbp]
    \centering
    \includegraphics[trim={1cm 6.5cm 0cm 0cm},clip,width=\linewidth]{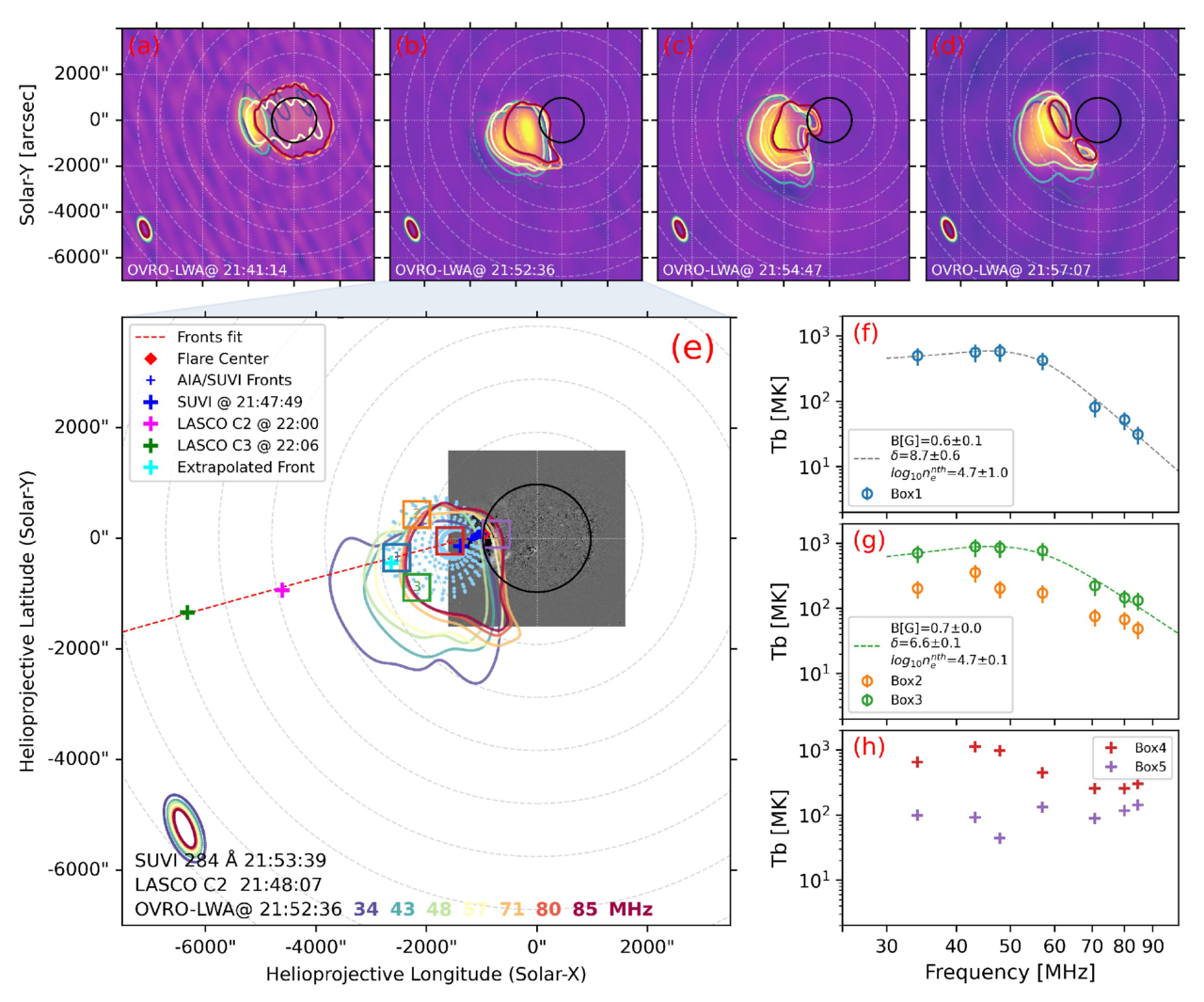}\\
    \includegraphics[trim={0.7cm 0.5cm 0cm 3cm},clip,width=0.93\linewidth]{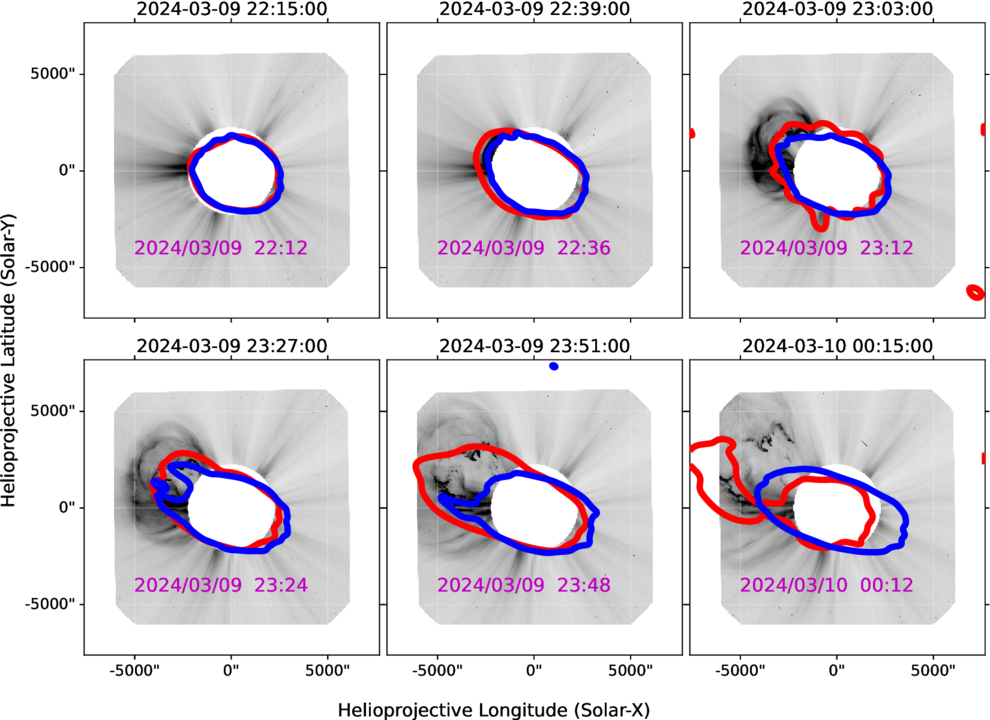}
    \caption{Top panels: Gyrosynchrotron emission detected from a CME on 31 December 2023 using OVRO-LWA at multiple timestamps during its propagation through the middle corona. The background in each panel shows the 48 MHz image, with overlaid contours at 34, 43, 48, 57, 71, 80, and 85 MHz, plotted at 10\% of the peak brightness temperature. Point spread functions for each frequency are shown in the bottom left of each panel. The black circle shows the solar disk. (Reproduced from \citet{chen2025measuring}). Bottom panels: The first possible gyroresonance emission detected from a CME event on 09 March 2024 using OVRO-LWA observations. Red and blue contours represent emission at 39 and 80 MHz. (Reproduced from \citet{mondal2025possible}).}
    \label{fig:overolwa_cme}
\end{figure}

Recent OVRO--LWA observations have demonstrated magnetic field measurements of a CME flux rope \citep{chen2025measuring} using gyrosynchrotron emission. The observation reveals a broad gyrosynchrotron source co-spatial with the CME leading edge at $\sim 1.8\,R_{\odot}$ as shown in the top panels of the Figure \ref{fig:overolwa_cme}, with a spectral peak near $\sim 50$~MHz. Spectral modeling indicates a magnetic field of $\approx 0.6$~G at this height, consistent with magnetic-flux conservation when compared with $\sim 300$~G fields in the low corona. These results show the observational capability of decametric observations using OVRO-LWA for measuring CME magnetic field in the middle corona.

\subsection{Detection of Thermal Gyroresonance Emission}
The high sensitivity of the SKA would also be instrumental in detecting the thermal emission from CMEs. Recently \citet{mondal2025a} presented the first evidence of thermal gyroresonance emission from a CME on 09 March 2024 observed using OVRO-LWA, shown in the bottom panels of Figure \ref{fig:overolwa_cme}. Thermal gyroresonance is only observable when the density of nonthermal electrons is not sufficient to emit detectable gyrosynchrotron emission. This has opened up a very interesting method to measure the CME magnetic field because thermal electrons and the magnetic field are always present in a CME, and hence thermal emission modulated by the CME magnetic field is also emitted by a CME. Modeling the thermal emission in the presence of the magnetic field would allow the determination of the magnetic field. This suggests that, irrespective of whether the CME has sufficient nonthermal electrons, it would be possible to measure the CME magnetic field by detailed modeling of the emission spectrum from the CME.

\subsection{Measuring CME Magnetic Field in Inner Heliosphere}\label{sec:current_status}
Faraday rotation measurements of background linearly polarised radio sources observed through the CME plasma provide a powerful probe of CME magnetic fields in the outer corona and inner heliosphere. The rotation of the polarization plane as radiation propagates through a magnetized plasma is quantified by the rotation measure (RM), given by $\mathrm{RM} = C_{\mathrm{FR}} \int n_e B_{\parallel}\, ds$, which contains the magnetic field information, and the resulting Faraday rotation (FR) scales as $\mathrm{FR} \propto \lambda^2 \mathrm{RM}$. This method is named ``heliopolarimetry". 

Previous heliopolarimetry efforts have primarily relied on high-frequency Jansky Very Large Array observations of a small number of polarized background sources \citep{kooi2017,Kooi2021}. However, the relatively narrow field-of-view ($\sim45$ arcmin at 1~GHz) and the use of higher observing frequencies restrict such measurements to a limited number of lines-of-sight and to comparatively low coronal heights ($<15\,R_{\odot}$). As a consequence, the sampling is sparse, and the reconstruction of CME magnetic field structures is subject to large uncertainties, particularly because it relies on the implicit assumption that the CME maintains its structural integrity as observations cycle through different line-of-sight. 

To make heliopolarimetry a transformative tool for space-weather research, observations must extend to higher coronal heights and the inner heliosphere, where decreasing magnetic fields and electron densities make low frequencies ($\sim 80$--$1000$~MHz) essential for detecting the small RM variations. Wide field-of-view, low-frequency instruments capable of simultaneously sampling many background sources across an expanding CME will therefore enable dense simultaneous line-of-sight sampling and time-resolved magnetic-field constraints—moving heliopolarimetry from sparse demonstrations toward routine space weather applications.

\subsection{Capabilities of SKA-Precursors and Pathfinders}\label{sec:new_instrument}
Over the last decade, several new-generation radio interferometers have become operational, which fulfill some basic requirements of heliopolarimetry, wide field-of-view, wide frequency coverage from low to high frequency, polarization purity of the instrument, and good sensitivity over small spectro-temporal integration. These telescopes include the SKA-precursors; MWA and MeerKAT, as well as SKA-pathfinder; LOw Frequency ARray \citep[LOFAR,][]{lofar2013}, Australian Square Kilometre Array Pathfinder \citep[ASKAP,][]{Hotan2021}. SKA telescopes, with their instantaneous broadband observing capabilities and higher sensitivity, will improve the instrumental RM precision significantly. Moreover, high-frequency bands of SKA-mid will allow us to perform heliopolarimetry observations at coronal heights as well, providing complementary constraints in conjunction with observations of direct emission from CME plasma as described in Section \ref{sec:radio_diag}.

Each of these instruments operates at different frequency ranges with different observing bandwidths. To measure the RM contribution from the CME, we need precision in the measurement of relative rotation measure (RRM = RM$_\mathrm{obs}$ - RM$_\mathrm{source}$) to be more than the RM of the CME. For a given instrument and frequency range, the maximum precision in the RRM is given as $\delta\text{RRM} \sim\frac{\delta\phi_{\text{FWHM}}}{\sqrt{2}\text{SNR}}$. $\delta\phi_{\text{FWHM}} = 2\sqrt{3}/\Delta\lambda^2$ is the half maximum of the full width of the rotation measure synthesis function \citep{Brentjens2005}, where $\Delta\lambda^2$ is the wavelength squared span across the observing bandwidth.   

\begin{figure*}[!htbp]
    \centering
    \includegraphics[trim={0.1cm 0.3cm 0.1cm 0.4cm},clip,scale=0.9]{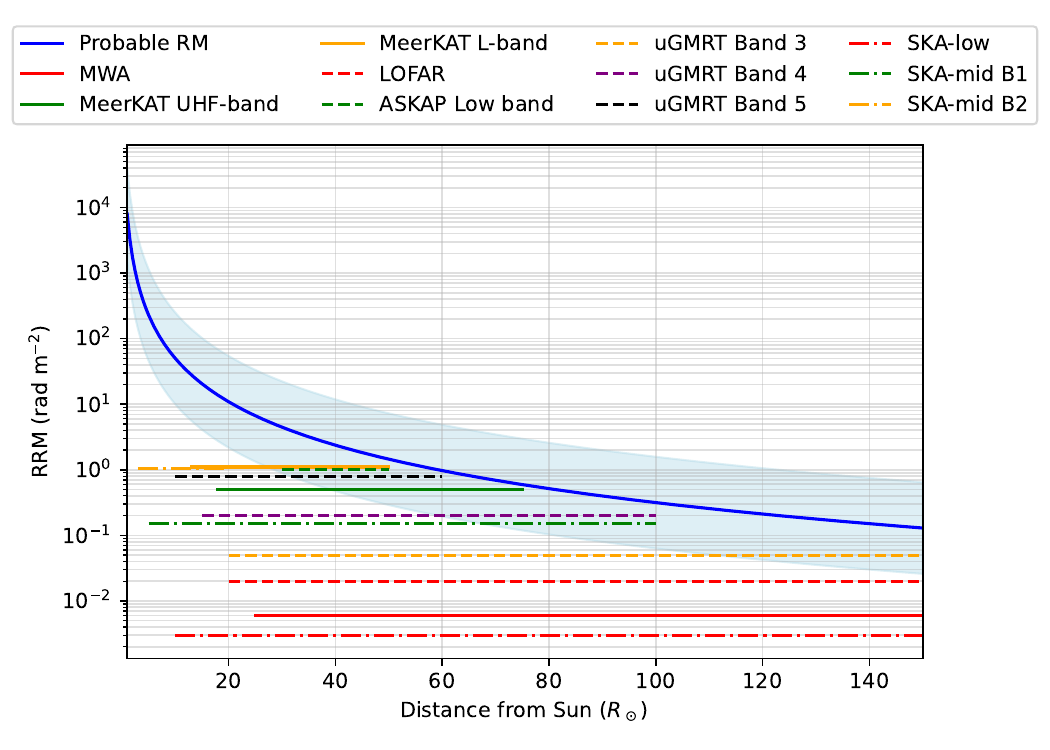}
    \caption{The solid blue curve shows the expected variation of relative rotation measure (RRM) of CMEs with heliocentric distance, with the light blue band indicating an order-of-magnitude uncertainty. Horizontal lines mark the achievable RRM precision for $\sim$15 minutes of integrations across different instruments and observing bands: solid for SKA precursors, dashed for SKA pathfinders, and dot–dashed for SKA bands. The horizontal extent of each line indicates the range of heliocentric distances accessible to that measurement, with an upper limit of 150 $R_\odot$ set by current ionospheric RM precision.
    }
    \label{fig:probable_RM}
    \vspace{-0.3cm}
\end{figure*}

Figure \ref{fig:probable_RM} shows the expected variation of CME-induced RRM with heliocentric distance \citep{Jensen2010,Oberoi2012}, with the shaded region indicating an order-of-magnitude uncertainty due to the lack of knowledge from direct measurements in the outer corona and inner heliosphere. Horizontal lines denote the RRM precision achievable with $\sim$15~minutes of integrations. These values represent the maximum heliocentric distances at which each instrument can detect CME-driven RRM variations. Existing facilities—including MWA, LOFAR, MeerKAT, and ASKAP—have already demonstrated excellent polarimetric performance and provide reference RM maps for detecting CME-induced variations. Wide field-of-view instruments can sample many background sources across a CME, but observations too close to the Sun increase system temperature and solar contamination, making calibration difficult and reducing sensitivity. Thus, each telescope has a minimum solar elongation constraint, which can be slightly different from the values mentioned in Table \ref{table:table1} depending on the instrumental primary beam response. 

\begin{table*}[!htbp]
\centering
    \renewcommand{\arraystretch}{1.4}
    \begin{tabular}{|p{3cm}|p{3.5cm}|p{2.2cm}|p{2.2cm}|}
    \hline
       Telescope Name & Frequency coverage & Minimum $R_\odot$ & Maximum $R_\odot$ \\ \hline \hline 
        MWA & 80 -- 300 MHz & 25 & 150\\
        \hline
        LOFAR & 110 -- 240 MHz & 20 & 150 \\
        \hline
        ASKAP-Low & 740 -- 1060 MHz & 30 & 50 \\
        \hline
        MeerKAT-UHF & 580 -- 1015 MHz & 18 & 75\\
       \hline
        MeerKAT-L & 900 -- 1670 MHz & 13 & 50\\
       \hline
        SKA-low & 50--350 MHz & 10 & 150\\
        \hline
        SKA-mid (Band-1) & 350--1050 MHz & 5 & 70\\
        \hline
        SKA-mid (Band-2) & 950--1760 MHz & 3 & 20 \\
        \hline
    \end{tabular}
    \caption{Frequency coverage, approximate minimum, and maximum solar elongation allowed by the instrumental RM precision for each of the instruments.}
    \label{table:table1}
    \vspace{-0.5cm}
\end{table*}

SKA-low and SKA-mid will increase the capability of detecting larger number of polarised sources substantially, enabling dense sampling on the sky, time-resolved heliopolarimetry to far greater heliocentric distances. Combining the inner limits with the maximum heliocentric distances from RRM sensitivity and broader frequency coverage of SKA-low and SKA-mid will cover heliocentric range $\sim3$--$150\,R_{\odot}$, a largely unexplored region of CME magnetic field measurements, that is crucial for understanding CME kinematics \citep[e.g.,][]{Colaninno_Vourlidas_Wu_2013, Sachdeva_etal_2017}. However, these observations require several hurdles to overcome. A detailed description of these requirements for heliopolarimetry measurements using the SKA telescopes is available in another chapter \citep{Oberoi01.2026.SKA}.

\subsection{Overcoming Challenges in Observations}\label{subsec:obs_challenges}
Although new-generation instruments can achieve the required sensitivity and polarimetric precision, heliopolarimetry faces two major challenges: capturing transient CME events and accurately separating ionospheric RM contributions from the measurements. Significant progress has been made on the first front, with successful event-based observations recently demonstrated using the MWA and ASKAP. We discussed these recent developments in Section \ref{subsec:heliopol_trigger}.

\subsubsection{Triggered Heliopolarimetry Observations using MWA and ASKAP}\label{subsec:heliopol_trigger}
Although CMEs typically originate from active regions and filaments, their exact onset time, propagation direction, and speed cannot be predicted with the precision needed for pre-scheduled observations. Blind monitoring near the Sun would consume substantial telescope time on already oversubscribed facilities, while yielding useful CME detections only rarely — and even then, the CME may lie outside the observed field-of-view. Therefore, efficient use of new-generation instruments, especially SKA telescopes with different key science commitments, requires an automated triggering system that can rapidly direct observations to the correct region of the sky at the right time to capture CMEs for heliopolarimetry.

\begin{figure}[!htbp]
    \centering
    \includegraphics[trim={0cm 4.5cm 0cm 0cm},clip,scale=0.6]{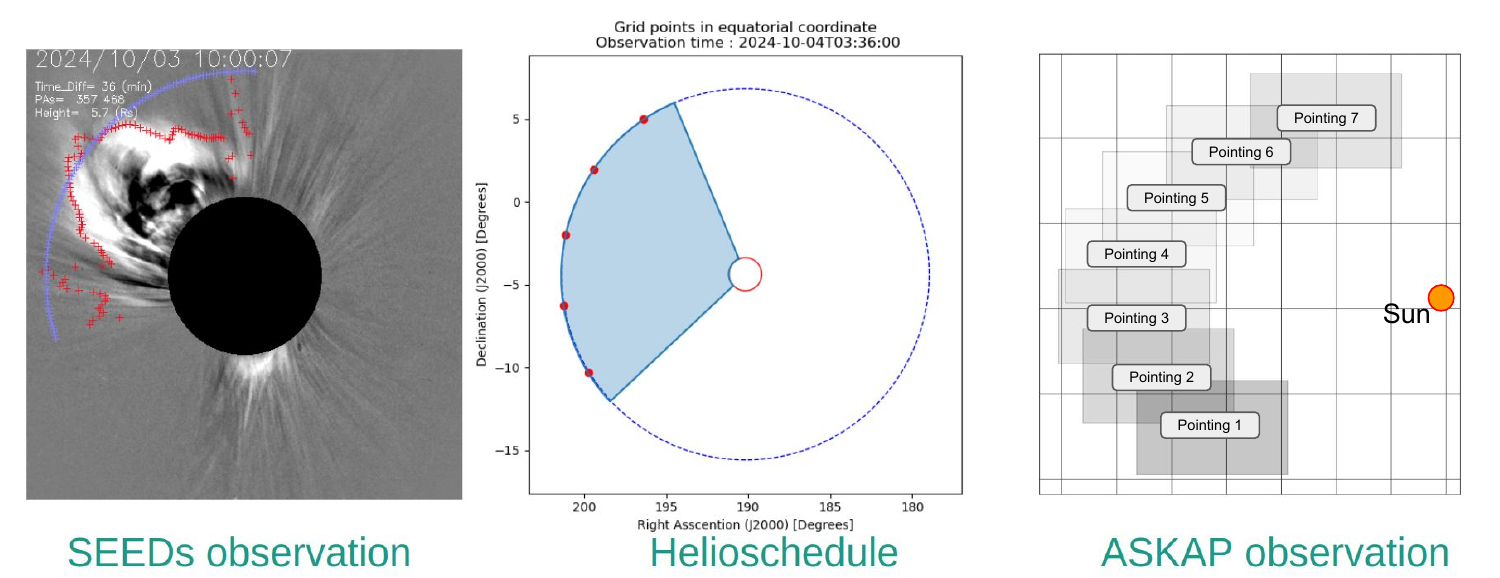}\\
    \includegraphics[trim={0cm 5cm 0cm 0cm},clip,scale=0.6]{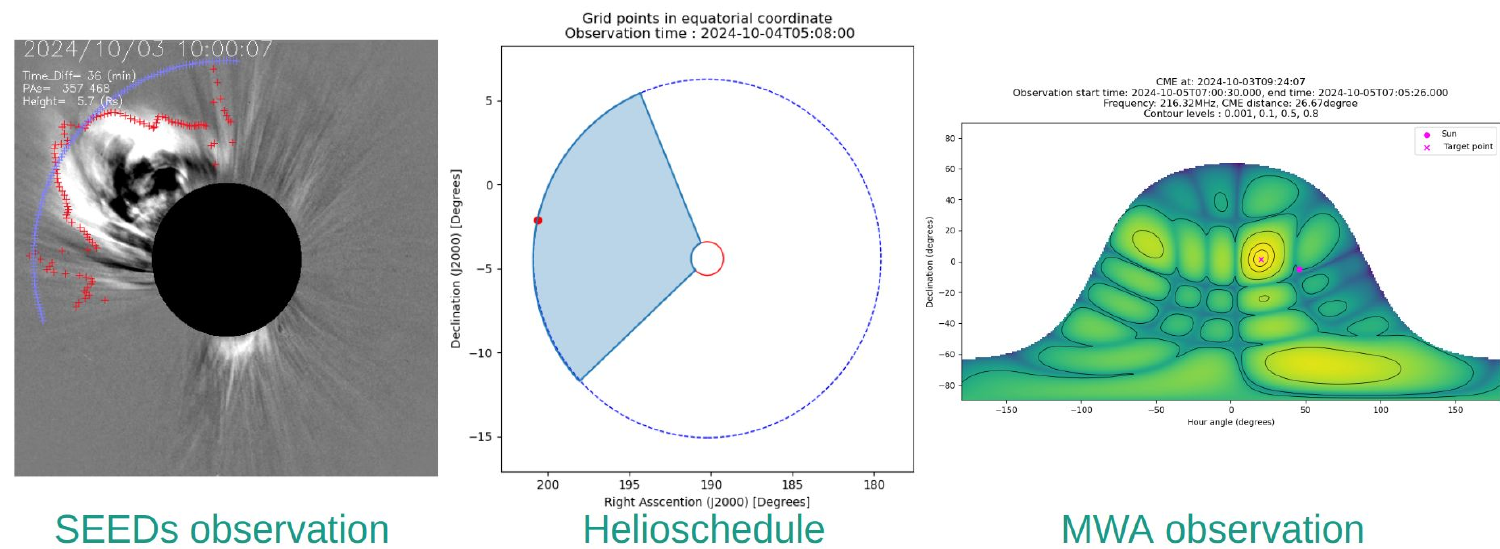}
    \caption{The top row shows three stages of \textsf{helioschedule} for ASKAP heliopolarimetry observation of a CME, and the bottom row shows the MWA observation of the same event, both triggered by \textsf{helioschedule}. The left panels illustrate CME detection by SEEDS using near-real-time coronagraph data, the middle panels display the calculated pointing configuration, and the right panels present the final telescope pointings used for the observations to keep the solar contamination minimum.}
    \label{fig:helioschedule}
\end{figure}

To address this observational challenge, an automated triggering system, \textsf{helioschedule}, has been developed, which directs the telescope to observe CMEs at the correct time and direction on the sky. The system relies on the near–real-time Solar Eruptive Event Detection System \citep[SEEDS;][]{SEEDS2008}, which identifies CMEs from white-light coronagraph data (shown in the left panels of Figure \ref{fig:helioschedule}) and updates their dynamical and geometrical properties approximately every hour. In practice, the time latency may range from one to five hours due to the availability of near-real-time data space-based coronagraphs constrained by telemetry availability. \textsf{Helioschedule} continuously monitors these updates and, when a CME exceeds user-defined thresholds in speed ($>200$ km/s) and angular width ($>60$deg), it initiates the triggering process. \textsf{Helioschedule} employs a drag-based model \citep[DBM;][]{DBM2021} to predict the CME arrival time at a specified heliocentric distance, using the CME speed estimated by SEEDS from coronagraph observations, and determine telescope pointings to observe the CME at that heliocentric height (shown in the middle panels of Figure \ref{fig:helioschedule}). The DBM accounts for solar-wind drag and typically yields arrival-time predictions with a median accuracy of $\sim$5~hours \citep{DBM2021}, which is adequate for targeting CMEs with wide field-of-view instruments.

For ASKAP, \textsf{helioschedule} selects the telescope pointings such that the Sun remains $\sim$11$^\circ$ off-axis of the telescope bore axis and adjacent fields overlap. Once these are defined, \textsf{helioschedule} forwards the pointing information and required metadata to the ASKAP scheduling system, Scheduling Autonomously Under Reactive Observational Needs \citep[SAURON;][]{sauron2022}. When the telescope is in the ASKAP Band 1 (800-1088 MHz) configuration -- optimal for RM sensitivity required for heliopolarimetry -- SAURON initiates observations of the selected fields at the predicted CME arrival time, typically spending $\sim$15~minutes in each pointing (shown in the top right panel of Figure \ref{fig:helioschedule}). The top panel of Figure \ref{fig:helioschedule} demonstrates steps in the \textsf{helioschedule} for one of the ASKAP observations. 

For the MWA, \textsf{helioschedule} defines the pointings such that the Sun lies near a beam null. A CME is tracked in steps of $\sim$20~$R_{\odot}$, and at each step, observations cycle through five contiguous 30.72~MHz sub-bands from 80 to 240 MHz. After determining scheduling time and pointings for the observation (shown in the bottom right panel of Figure \ref{fig:helioschedule}), \textsf{helioschedule} employs the existing web-based triggering interface \citep{Hancock} to command the MWA to observe the prescribed pointings at the designated times. The bottom panel of Figure \ref{fig:helioschedule} demonstrates one such MWA observation done using the \textsf{helioschedule}.

\textsf{Helioschedule} effectively overcomes the major observational challenge of blindly observing transient CMEs, enabling efficient use of valuable telescope time for heliopolarimetry. The system has been operational on both the MWA and ASKAP since August 2024, and has successfully triggered observations of $\sim$60 CMEs with the MWA and $\sim$15 CMEs with ASKAP to date.

\subsubsection{Removal of Ionospheric RM}
\begin{figure*}[!htbp]
    \centering
    \includegraphics[width=0.6\linewidth]{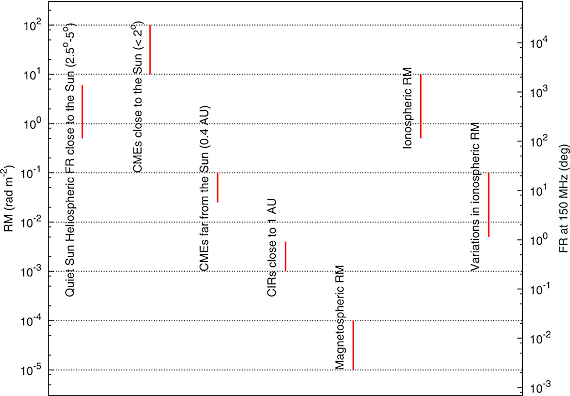}
    \caption{Rotation measure contribution from different heliospheric and ionospheric media. (Reproduced from \citet{oberoi2012_rms}).}
    \label{fig:rm_compare}
\end{figure*}

The second major hurdle for low-frequency observations targeting the larger solar elongations is the contribution of the ionosphere to the measured RM. Figure \ref{fig:rm_compare} shows the contribution of RMs from heliospheric and ionospheric plasma \citep{oberoi2012_rms}. Currently available and routinely used ionospheric RM correction tools, like \textsf{ionFR} \citep{Sotomayor2013} and \textsf{RMextract} \citep{Gasperin2018}, can routinely provide ionospheric RM at a precision of $\sim0.1-0.3\ \mathrm{rad}\ m^{-2}$. This accuracy is sufficient to measure RM due to the CME up to $\sim$60 $R_\odot$ (Figure \ref{fig:probable_RM}). At much larger distances ($>60\ R_\odot$), the RM contribution by the CME is comparable to the variability of ionospheric RM \citep{oberoi2012_rms}. This requires ionospheric RM estimation at a precision of $\sim0.01\ \mathrm{rad}\ m^{-2}$ with higher time resolution compared to the currently available measurements using the Global Navigation Satellite System. Efforts are underway to develop to meet the required RM-precision of $\sim0.01\ \mathrm{rad}\ m^{-2}$ using a tri-band CubeSat \citep{LANTER20235503}, which could provide direct measurements of ionospheric RM. 

\section{Role of SKA for Advancing CME Magnetic Field Measurements}\label{sec:ska_role}
SKA precursors and pathfinders have already introduced several promising methods for probing CME magnetic fields in both the corona and heliosphere. Building on this progress, the greatly improved sensitivity of SKA, broad instantaneous bandwidth, superior surface-brightness sensitivity from enhanced UV coverage, and its extensive overall frequency range will together enable transformative advances in CME magnetic field measurements and in our understanding of CME physics. Realizing this potential, however, will require robust calibration, sophisticated data processing, and well-designed observing strategies, which are described in detail in another chapter \citep{Oberoi01.2026.SKA}. In the following section, we outline how the SKA will address current limitations and advance the field.

\subsection{Robust modeling of CME Gyrosynchrotron Emission}
SKA precursors and pathfinders have demonstrated that faint GS emission from CME plasma can be detected out to mid-coronal heights, even for relatively weak CMEs. Because GS models depend on multiple CME plasma parameters that can be degenerate, spectropolarimetric imaging is crucial for breaking these degeneracies and accounting for inhomogeneous plasma structure \citep{Kansabanik_2024}. Previous studies \citep{Kansabanik2023_CME1, Kansabanik_2024} have incorporated polarimetric upper limits into GS modeling, with constraints at the level of about 5–10\% of Stokes I.

The much higher sensitivity of the SKA will enable routine detection of the weak circularly polarized component of CME GS emission. At larger heliocentric heights, observations often miss the spectral peak, introducing significant uncertainties in the derived GS parameters. SKA-Low’s broad frequency coverage will allow sampling across a wide range of heights -- including the spectral peak -- thereby enabling accurate magnetic field estimates. SKA-mid will complement this by probing lower coronal heights, particularly during the eruption phase and in regions of magnetic reconnection, offering new insights into CME initiation and early magnetic evolution.

Without advance knowledge of eruptions, CME GS observations require either continuous daytime monitoring or very fast triggering. A dedicated daytime SKA-low station beam could provide uninterrupted coverage, with only intervals containing coronagraph-identified CMEs saved and processed to minimize data volume. Triggering is appealing, but current space-based coronagraphs introduce delays of several hours—too slow for effective early phase observation and follow-up in the lower and middle corona. The forthcoming SunCET \citep{Mason2021} EUV cubesat mission will transmit rapid CME alerts, dramatically cutting this latency. Integrating SunCET CME beacons with a robust triggering system will be key to efficient use of SKA observing time and to capturing a large fraction of CME events with both SKA telescopes.

\subsection{A New Observing Tool using Thermal Emission}
Gyrosynchrotron (GS) emission requires the presence of mildly relativistic electrons, whereas thermal electrons are ubiquitous within CME plasma. As thermal free–free emission propagates through a magnetized medium, it can acquire circular polarization \citep{Sastry_2009}, although the expected polarization fraction is typically $\lesssim 1\%$. For a given magnetic field strength, the polarisation fraction increases toward lower frequencies. Despite this, circularly polarised thermal free–free emission from CME plasma has not yet been observationally confirmed. The enhanced sensitivity, broad frequency coverage, and improved spatial resolution of SKA-Low at meter wavelengths are expected to enable the detection of this weak polarisation signal, thereby opening a new avenue for direct measurements of CME magnetic fields in the middle corona.

Another thermal emission mechanism capable of providing magnetic field diagnostics is gyroresonance emission, which has recently been detected from CME plasma (bottom panel of Figure \ref{fig:overolwa_cme}) using OVRO–LWA \citep{mondal2025a}. This discovery highlights an additional promising technique for probing CME magnetic fields with the SKA. Gyroresonance emission is characterized by strong circular polarization at low harmonics of the gyrofrequency, providing a direct estimation of CME magnetic field strength. This makes the SKA-Low band particularly well suited to observing the second and third harmonics corresponding to typical CME magnetic field strengths of $\sim$1 G in the middle corona.

\subsection{Robust modeling using Multi Line-of-sight FR Measurements}
Modeling CME magnetic fields in the outer corona and inner heliosphere using FR observations requires simultaneous measurements along multiple lines-of-sight. However, this requires a triggering system to observe the right patch of the sky at the right time to observe the background sources through the CME plasma. SKA precursors and pathfinders have shown that this is feasible, as discussed in Section \ref{subsec:heliopol_trigger}, and likewise need a trigger-based observing strategy to optimally capture heliospheric CMEs using the SKA telescopes.

Current polarized source densities reachable in $\sim$15 minutes are about 0.1–0.5 sources deg$^{-2}$ at SKA-low and $\sim$1 source deg$^{-2}$ at SKA-mid bands using precursors and pathfinder telescopes. Although this is a significant advancement from earlier single-source observations at a single pointing using narrow field-of-view instruments like the JVLA, higher source density will certainly improve the reconstruction of the 3D CME magnetic field from line-of-sight measurements. With the higher sensitivity of SKA telescopes and reduced beam depolarization from improved angular resolution, these densities will rise significantly, strengthening constraints on CME magnetic field models.

In addition, the broader fractional bandwidth of SKA-low will provide greater RM sensitivity, enabling the detection of weaker CME magnetic fields farther from the Sun. Joint observations using both SKA telescopes will extend coverage from mid-coronal heights into the inner heliosphere. Together, these advances will make FR observations a powerful tool for accurately modeling CME magnetic fields and their spatial and temporal evolution, ultimately improving predictions of CME geoeffectiveness.

\section{Conclusion}\label{sec:conclusion}
Remote sensing of the CME magnetic field remains one of the most long-standing and formidable challenges in solar and heliospheric physics. Despite decades of CME observations using white-light coronagraphs and heliospheric imagers, as well as the development of increasingly sophisticated modeling frameworks, the accuracy of SpWx predictions remains limited. This shortcoming primarily arises from the lack of routine, spatially resolved magnetic field measurements of the full CME structure in the corona and heliosphere—measurements that single or limited-vantage in-situ observations cannot provide. Consequently, remote sensing of CME magnetic fields represents the only viable path forward for achieving transformative improvements in CME magnetic field characterization.

Although radio observations have long promised several unique remote-sensing techniques for measuring CME magnetic fields from the ground \citep{Vourlidas2020, Carley2020}, these approaches have not yet entered mainstream use for SpWx prediction. This is due to a combination of factors, including the historical lack of sufficiently capable radio instruments, limited computational resources, inadequate human capacity, and the absence of mature modeling frameworks and analysis tools required to fully exploit such observations.

SKA precursors and pathfinder instruments have demonstrated the feasibility of these radio diagnostics and, in some cases, enabled such observations for the first time. While these results highlight the scientific potential of radio-based CME magnetic field measurements, they have also revealed key limitations inherent to the precursor and pathfinder facilities. These limitations are primarily associated with insufficient sensitivity, limited imaging fidelity, restricted frequency coverage, and narrow instantaneous bandwidth. The SKA telescopes will provide substantial improvements in all of these areas and are therefore expected to overcome the current observational constraints, enabling robust radio measurements of CME magnetic fields, which will be routinely available.

In parallel with advances in observational capabilities, the community must place equal emphasis on the development of modeling frameworks. At present, most CME and space-weather models are typically validated a posteriori using radio observations, but they do not directly assimilate radio observables as model inputs or constraints. With recent advances in computational capabilities, the availability of efficient and scalable algorithms, and the rapid growth of machine-learning and artificial-intelligence–based techniques, there is now a timely opportunity to incorporate radio observational constraints into CME and space-weather models. To fully exploit the unprecedented sensitivity and fidelity of SKA observations for space-weather research and prediction, sustained effort in this parallel domain—spanning modeling, data assimilation, and analysis tools—will be essential.

\section{Acknowledgment}
This work uses observations from the MWA from Inyarrimanha Ilgari Bundara, the CSIRO Murchison Radio-astronomy Observatory. We acknowledge the Wajarri Yamaji people as the traditional owners and native title holders of the observatory site. Support for the operation of the MWA is provided by the Australian Government (NCRIS), under a contract to Curtin University administered by Astronomy Australia Limited. This work also uses observations from the MeerKAT radio telescope. The MeerKAT telescope is operated by the South African Radio Astronomy Observatory, which is a facility of the National Research Foundation, an agency of the Department of Science, Technology, and Innovation. The OVRO-LWA expansion project was supported by NSF under grant AST-1828784. OVRO-LWA operations for solar and space weather sciences are supported by the National Science Foundation (NSF) under grant AGS-2436999. D.K. acknowledges financial support from the Severo Ochoa grant CEX2021-001131-S and from the Spanish grant PID2023-147883NB-C21, funded by MCIU/AEI/ 10.13039/501100011033, as well as support through ERDF/EU, and the NASA Living with a Star Jack Eddy Postdoctoral Fellowship Program, administered by UCAR’s Cooperative Programs for the Advancement of Earth System Science (CPAESS) under award 80NSSC22M0097. P.M. and D.O. acknowledge support from the Department of Atomic Energy, Government of India, under the project no. 12-R\&D-TFR-5.02-0700. A.V. was supported by NASA grants 80NSSC21K1860 and 80NSSC22K0970. S.P. would like to acknowledge support from the CEFIPRA Research Project No. 6904-2.

\bibliographystyle{abbrvnat-maxbibnames4}
\bibliography{chapter} 

\end{document}